\documentclass[twocolumn,showpacs,amsmath,amssymb,nofootinbib]{revtex4-1}
\usepackage{dcolumn}
\usepackage{bm}
\usepackage{graphicx}
\usepackage{color}

\newcommand{\be}{\begin{equation}}
\newcommand{\ee}{\end{equation}}
\newcommand{\ba}{\begin{eqnarray}}
\newcommand{\ea}{\end{eqnarray}}
\newcommand{\non}{\nonumber}
\newcommand{\n}[1]{\label{#1}}
\newcommand{\eq}[1]{(\ref{#1})}
\newcommand{\hh}{\, ,\hspace{0.5cm}}
\newcommand{\hhh}{\, ,\hspace{0.1cm}}

\newcommand{\bi}[1]{\bibitem{#1}}

\begin{document}

\title{The {\em Universal Area Product}: An Heuristic Argument}
\author{Don N. Page}
\email{profdonpage@gmail.com}
\author{Andrey A. Shoom}
\email{ashoom@ualberta.ca}
\affiliation{Theoretical Physics Institute, University of Alberta,
Edmonton, AB, Canada,  T6G 2E1}
\date{2015 May 5, revised 2015 July 31}

\begin{abstract}
We present an heuristic argument for the {\em universal area product:} $A_{+}A_{-}=(8\pi J)^{2}+(4\pi Q^{2})^{2}$ for a four-dimensional, stationary, axisymmetric, electrically charged black hole with an arbitrary stationary axisymmetric distribution of external matter (possibly charged), derived by Marcus Ansorg and J\"org Hennig. Here $A_{+}$ and $A_{-}$ are the areas of the event and Cauchy horizons, and $J$ and $Q$ are the angular momentum and electric charge. Based on this argument, we conjecture that a universal area product holds for higher-dimensional, stationary, multi-horizon black objects in the presence of an external stationary charged distribution of matter.
\end{abstract}

\pacs{04.20.Cv, 04.70.Bw, 04.40.Nr \hfill Alberta-Thy-4-15}

\maketitle

The product of all horizon areas for general rotating multi-charge black objects in four- and higher-dimensional asymptotically flat or anti-de Sitter spacetimes depends only on the quantized charges, quantized angular momenta, and the cosmological constant \cite{CGP}. The remarkable result of quantization of the product of the horizon areas was already implicitly presented in the earlier works by Larsen and Cveti\v c \cite{Lar1,CvetLar1,CvetLar2,CvetLar3,Lar2} where the entropies of the horizons (proportional to their areas) of four- and five-dimensional black holes were expressed in terms of the excitation numbers of the left and right moving modes of a weakly-coupled two-dimensional conformal field theory. The quantization rule of the product of the areas follows from this rule \cite{CvetLar4}.  Recently Goldstein, Jejjala, and Nampuri \cite{Goldstein:2014gta} showed that the product of areas for fixed $U(1)$ charges is also an invariant for all static spherically symmetric black holes in four-dimensional $N = 2$ supergravity.

As it was stated by Cveti\v c, Gibbons, and Pope \cite{CGP}, if the cosmological constant is quantized, the area products may provide a ``looking glass'' for probing the microscopics of black objects. The solutions considered in \cite{CGP} represent isolated black objects. Thus, it is important to study whether the area products survive in the presence of external matter and fields. 

Here we present arguments for a conjecture that this is likely to be the case. Our conjecture is based on a heuristic argument for the universal area product of the horizon areas of a four-dimensional stationary and electrically charged black hole distorted by arbitrary stationary axisymmetric electromagnetic and gravitational fields which could be due to an electrically charged stationary distribution of matter around the black hole.

The relation\footnote{We use a system of units with $G=c=1$ and spacetime signature $(+ - - -)$.} $A^{2}=(8\pi J)^{2}+(4\pi Q^{2})^{2}$, where $A$ is the extremal horizon area, $J$ is the angular momentum, and $Q$ is the electric charge, was proven for an electrically charged, extremal black hole with arbitrary surrounding matter by Ansorg and Pfister \cite{AP}. A universal area product of the horizon areas of a stationary and axisymmetric black hole surrounded by matter, $A_{+}A_{-}=(8\pi J)^{2}$, where $A_{+}$ and $A_{-}$ are the areas of the event horizon (EH) and Cauchy horizon (CH), was constructed by Ansorg and Hennig in \cite{J1}, and later its validity was numerically illustrated in \cite{J2}. Finally, the universal area product $A_{+}A_{-}=(8\pi J)^{2}+(4\pi Q^{2})^{2}$ for a four-dimensional, stationary, axisymmetric, electrically charged black hole with surrounding matter was proved by Ansorg and Hennig \cite{AH1,AH2}. One can check directly that such a relation holds for a Kerr-Newman black hole (see, e.g., \cite{MTW}). 

It is rather surprising that the same relation holds in the presence of distortion of the black hole by, for example, surrounding matter. Ansorg and Hennig \cite{AH1,AH2} proved the validity of this relation by using the inverse scattering method applied to the linear matrix problem whose integrability conditions are equivalent to two complex Ernst equations representing the Einstein-Maxwell system. Here we give an heuristic argument explaining such a relation. 

Our model consists of a distorted Kerr-Newman black hole of angular momentum $J$, electric charge $Q$, and a surrounding stationary, axisymmetric, and, in general, electrically charged distribution of matter representing distortion sources which is defined by a set of parameters, $D_{i}$\footnote{A solution representing a  distorted Kerr-Newman black hole in an external static gravitational field was constructed by Breton, Garcia, Manko, and Denisova \cite{BGMD}.}. The space-time beyond the sources is asymptotically flat, and the black hole is in stationary equilibrium. 

Let us consider a quasi-stationary transition from an undistorted Kerr-Newman black hole solution to another stationary and axisymmetric but distorted black hole defined by different values of the parameters $D_{i}$, such that the black hole remains close to equilibrium during the transition. An undistorted black hole corresponds to vanishing $D_{i}$, e.g. when the distortion sources are located at infinity. The quasi-stationary transition consists of small transition steps, such that at the end of each step the black hole settles down to another stationary state. During each step we slowly bring the sources towards the black hole in such a way that the space-time axial symmetry is preserved. 

In such a case, the generated gravitational and electromagnetic waves are weak and do not carry angular momentum. These waves accompany a transition from one value of the distortion field to the next that is infinitesimally nearby.  Such a transition can be considered as a perturbation of the initial state defined by the metric $g^{(i)}_{\alpha\beta}$ and the electromagnetic 4-vector potential $A^{(i)}_{\alpha}$. The result of one transition step is the final state $g^{(f)}_{\alpha\beta}$ and $A^{(f)}_{\alpha}$ which can be expressed as follows:
\ba
g^{(f)}_{\alpha\beta}&=&g^{(i)}_{\alpha\beta}+\lambda h_{\alpha\beta}+O(\lambda^{2},\lambda\mu,\mu^{2})\,,\n{1a}\\
A^{(f)}_{\alpha}&=&A^{(i)}_{\alpha}+\mu B_{\alpha}+O(\lambda^{2},\lambda\mu,\mu^{2})\,,\n{1b}
\ea
where at the black hole horizon $h_{\alpha\beta}=O(g_{\alpha\beta})$ and $g^{\alpha\beta}B_{\alpha}B_{\beta}=O(1)$, and where $\lambda,\mu\ll1$ are dimensionless parameters defining the strength of the gravitational and electromagnetic perturbations, respectively. 

The perturbation during one transition step grows from zero to the final value corresponding to the new distortion field during a finite time interval. After the perturbation reaches its final value, the gravitational and electromagnetic waves decay as an inverse power of the advanced Eddington coordinate $v$ \cite{Ori2,MarOri}, and the black hole settles down to another state defined by new values of the parameters $D_{i}$. Note that according to the nature of the perturbation, the black hole's electric charge $Q$ and angular momentum $J$ remain fixed. We are assuming classical positive energy conditions that do not allow the black hole event horizon area $A_{+}$ to decrease, so for transitions that are adiabatic, $A_{+}$ must remain unchanged.

We shall show now that a sufficiently-slow quasi-stationary transition is indeed adiabatic, i.e. $A_{+}$ is an adiabatic invariant with respect to the perturbation. In order to do so, we shall follow the approach given by Hawking and Hartle \cite{HH}. Namely, we consider the Hawking-Hartle null tetrad $(l^{\alpha}, n^{\alpha}, m^{\alpha}, \bar{m}^{\alpha})$ which is well-behaved on the future EH (see, e.g., \cite{Chandra}). At the horizon, the null vector $l^{\alpha}=dx^{\alpha}/dt$ is a null geodesic generator of the null EH hypersurface. Here `$t$' is a non-affine parameter along the null geodesic generators of the EH, which we shall choose as one of the spacetime coordinates, so that $l^{\alpha}=\delta^{\alpha}_{t}$. 

At the fixed $t$ coordinate the EH surface is topologically a sphere on which the orbits of the Killing vector are closed circles (except for two fixed points on the symmetry axes). We choose $\phi$ to be the azimuthal Killing coordinate. Then, one can always choose the other coordinate $\theta$ on the sphere to be orthogonal to the $\phi$ coordinate.  

We choose $x$ as a coordinate which is constant ($x=0$) on the EH and orthogonal to the $\theta$ and $\phi$ coordinates. The complex null vector $m^{\alpha}$ and its complex conjugate $\bar{m}^{\alpha}$ for $x=0$ lie on the EH. There is gauge freedom in their spatial and null rotations. We fix the null rotation gauge by imposing $m^{t}=0$. To fix the spatial rotation we take $m^{\theta}$ real and $m^{\phi}$ imaginary. Given these null vectors, the null vector $n^{\alpha}$ is then uniquely defined through the null tetrad orthogonality conditions in which the only nonzero dot products are $l^{\alpha}n_{\alpha}=-m^{\alpha}\bar{m}_{\alpha}=1$. In the given coordinates we have the null tetrad on the EH,
\ba
&&l^{\alpha}=[1,0,0,0]\hhh n^{\alpha}=[n^{t},n^{x},0,0]\,,\non\\
&&m^{\alpha}=[0,0,e^{a},ie^{b}]\hhh \bar{m}^{\alpha}=[0,0,e^{a},-ie^{b}]\,,\n{2}
\ea
where the vector components are real functions of the coordinates $t$ and $\theta$ on the horizon. 

In this null tetrad the first two of the Newman-Penrose equations (see \cite{Chandra,NP}) take the following form at the black hole EH:
\ba
\frac{d\rho}{dt}&=&\rho^{2}+\sigma\bar{\sigma}+2\epsilon\rho+\Phi_{00}\n{3a}\,,\\
\frac{d\sigma}{dt}&=&2\rho\sigma+2\epsilon\sigma+\Psi_{0}\n{3b}\,,
\ea
where $\rho=l_{\beta;\alpha}m^{\alpha}\bar{m}^{\beta}$ is real (because $l^{\alpha}$ is hypersurface-orthogonal) and measures the convergence of the null geodesic generators, and $\sigma=l_{\alpha;\beta}m^{\alpha}m^{\beta}$ is the shear, which is complex. For the tetrad \eq{2}, $\epsilon$ is real, and because $l^{\alpha}$ is a null geodesic generator, one has $l^{\alpha}l_{\beta;\alpha}=2\epsilon l_{\beta}$, which fixes the scaling of $l^{\alpha}$ on the EH in terms of $\epsilon$. For the null tetrad \eq{2} these spin coefficients read
\be
\rho=\frac{1}{2}(a_{,t}+b_{,t})\hh\sigma=\frac{1}{2}(a_{,t}-b_{,t})\hh\epsilon=-\frac{n^{x}_{,t}}{2n^{x}}\,.\n{3c}\
\ee

The complex Weyl scalar $\Psi_{0}=C_{\alpha\beta\gamma\delta}l^{\alpha}m^{\beta}l^{\gamma}m^{\delta}$ corresponds to an ingoing transverse gravitational wave. The transverse directions of the wave defined by the vectors $m^{\alpha}$ and $\bar{m}^{\alpha}$ are along the black hole horizon surface, and the wave propagates in the direction defined by the ingoing null vector $n^{\alpha}$. Using the expressions \eq{3b} and \eq{3c} we derive
\be\n{3d}
\Psi_{0}=\frac{1}{2}(a_{,tt}-b_{,tt}-a_{,t}^{2}+b_{,t}^{2})-\epsilon(a_{,t}-b_{,t})\,.
\ee

An ingoing electromagnetic wave is defined by the complex Ricci tensor component $\Phi_{00}=2\bar{\phi}_{0}\phi_{0}=-(1/2)R_{\alpha\beta}l^{\alpha}l^{\beta}$, where $\phi_{0}=F_{\alpha\beta}l^{\alpha}m^{\beta}$ is the Maxwell scalar. In the radiation gauge $A_{t}=0$, 
\be\n{3e}
\phi_{0}=A_{\theta,t}e^{a}+iA_{\phi,t}e^{b}\hhh \Phi_{00}=2(A_{\theta,t}^{2}e^{2a}+A_{\phi,t}^{2}e^{2b})\,.
\ee

Given the quantities above, one can calculate the rate of change of the EH area (cf. \cite{Haw}),
\be\n{4}
\frac{dA_{+}}{dt}=-2\int\rho\,dA_{+}\,,
\ee
where here and henceforth an integral over $dA_{+}$ denotes an integral over the closed two-dimensional surface of the EH at time $t$. Using the transition expressions \eq{1a}-\eq{1b} we can solve the Newman-Penrose equations perturbatively and find $\rho$, in order to calculate the EH area change. 

To construct the gravitational and electromagnetic perturbative expansions we observe that the gravitational perturbation $\Psi_{0}$ is of the first order in $\lambda$ and the electromagnetic perturbation $\phi_{0}$ is of the first order in $\mu$. The gravitational perturbation induces through its nonlinear interaction with the background electromagnetic field the term in $\phi_{0}$ which is of the first order in $\lambda$. As a result, the corresponding Ricci tensor component $\Phi_{00}$ is a quadratic expression in $\lambda$ and $\mu$. These terms induce terms of the corresponding order in $\Psi_{0}$, and so we get the following power series expansions: 
\ba
\Psi_{0}&=&\lambda \Psi^{(1,0)}_{0}+O(\lambda^{2},\lambda\mu, \mu^{2})\,,\n{5a}\\
\phi_{0}&=&\mu\phi^{(0,1)}_{0}+\lambda\phi_{0}^{(1,0)}+O(\mu^{2},\lambda\mu,\lambda^{2})\,,\n{5b}\\
\Phi_{00}&=&\mu^{2}\Phi^{(0,2)}_{00}+\lambda\mu\Phi^{(1,1)}_{00}+\lambda^{2}\Phi_{00}^{(2,0)}\non\\
&+&O(\mu^{3},\lambda\mu^{2},\lambda^{2}\mu,\lambda^{3})\,.\n{5c}
\ea

The corresponding expansions of the spin coefficients can be deduced from the Newman-Penrose equations. According to the order of $\Psi_{0}$ and $\Phi_{00}$, the leading terms of the convergence are quadratic in $\lambda$ and $\mu$, while the leading term of the shear is linear in $\lambda$, 
\ba
\rho&=&\lambda^{2}\rho^{(2,0)}+\lambda\mu\rho^{(1,1)}+\mu^{2}\rho^{(0,2)}\non\\
&+&O(\mu^{3},\lambda\mu^{2},\lambda^{2}\mu,\lambda^{3})\,,\n{5d}\\
\sigma&=&\lambda \sigma^{(1,0)}+O(\lambda^{2},\lambda\mu,\mu^{2})\,,\n{5e}\\
\epsilon&=&\epsilon^{(0)}+\lambda\epsilon^{(1,0)}+\mu\epsilon^{(0,1)}+O(\lambda^{2},\lambda\mu,\mu^{2})\,.\n{5f}
\ea
One can show that according to the equation $(l^{\alpha}l_{\alpha})_{;\beta}=-2\kappa l_{\beta}$ defining the surface gravity $\kappa$ when the metric is stationary, we have $2\epsilon^{(0)}=\kappa$. 

Substituting the expansions \eq{5a}-\eq{5f} into the Newman-Penrose equations and solving them we derive
\ba
\sigma^{(1,0)}(t)&=&-\int_{t}^{\infty}e^{-\kappa(t'-t)}\Psi_{0}^{(1,0)}(t')dt'\,,\n{6a}\\
\rho^{(2,0)}(t)&=&-\int_{t}^{\infty}e^{-\kappa(t'-t)}\left(|\sigma^{(1,0)}(t')|^{2}+\Phi^{(2,0)}_{00}(t')\right)dt'\,,
\non\\
\n{6b}\\
\rho^{(1,1)}(t)&=&-\int_{t}^{\infty}e^{-\kappa(t'-t)}\Phi^{(1,1)}_{00}(t')dt'\,,\n{6c}\\
\rho^{(0,2)}(t)&=&-\int_{t}^{\infty}e^{-\kappa(t'-t)}\Phi^{(0,2)}_{00}(t')dt'\,.\n{6d}
\ea
Here the convergence $\rho_{0}^{(2,0)}$ is due to the primary gravitational and induced electromagnetic waves, the convergence $\rho_{0}^{(1,1)}$ is due to the primary and induced electromagnetic waves, and the convergence $\rho^{(0,2)}$ is due to the primary electromagnetic wave. These convergences give the corresponding rate of change in the EH surface area.

According to our model, for one transition step the gravitational and electromagnetic perturbations begin at $t=t_{i}$ and end at $t=t_{f}$. Note that for $t<t_{i}$, $\sigma$ and $\rho$ are non-zero but exponentially small. This is because the solution \eq{6a}-\eq{6d} is defined by the entire future history of the black hole, what is a manifestation of the teleological nature of the EH. As a result, for the effect of one transition step we derive
\ba
\frac{dA_{+}}{dt}&=&-2\int\rho^{(net)}(t)\,dA_{+}\,,\n{7a}\\
\delta A_{+}&=&-2\int_{-\infty}^{+\infty}dt\int\rho^{(net)}(t)\,dA_{+}\,,\n{7b}\\
\rho^{(net)}(t)&=&\lambda^{2}\rho^{(2,0)}(t)+\lambda\mu\rho^{(1,1)}(t)+\mu^{2}\rho^{(0,2)}(t)\,.\n{7c}
\ea
The total quasi-stationary transition consists of $N$ such steps. 

Let $\Delta_{G}=|g_{(0)}^{\alpha\beta}||\tilde{g}_{\alpha\beta}-g^{(0)}_{\alpha\beta}|$ and $\Delta_{EM}=\sqrt{|g_{(0)}^{\alpha\beta}(\tilde{A}_{\alpha}-A^{(0)}_{\alpha})(\tilde{A}_{\beta}-A^{(0)}_{\beta})|}$ be the total dimensionless perturbations of the original metric $g^{(0)}_{\alpha\beta}$ and of the electromagnetic potential $A^{(0)}_{\alpha}$ representing an undistorted Kerr-Newman black hole which in a quasi-stationary transition is brought to the distorted metric $\tilde{g}_{\alpha\beta}$ and the corresponding electromagnetic potential $\tilde{A}_{\alpha}$. Then one gets $\lambda\sim\Delta_{G}/N$, $\mu\sim\Delta_{EM}/N$, and the total time of the quasi-stationary transition is of the order of $N\delta t$, where $\delta t=t_{f}-t_{i}$. 

To get an estimate of the total area change $\Delta A_{+}$ we model the perturbations $\Psi_{0}$ and $\Phi_{00}$ by rectangular impulses. One can define an upper bound crude estimate for the amplitudes of the perturbations $\Psi_{0}$ and $\Phi_{00}$. It follows from Eqs.\ \eq{3d} and \eq{3e}, and from the order of the surface gravity $\kappa=2\epsilon^{(0)}$, that on the EH the order of the amplitude of $\Psi_{0}$ is $1/(\delta t)^{2}$ for a fast transition, $\kappa\delta t\ll 1$, and is $\kappa/\delta t$ for a slow transition, $\kappa\delta t\gg 1$, while the order of magnitude of $\Phi_{00}$ is $1/(\delta t)^{2}$, regardless of the transition rate. 

One can check that in order to have the first terms on the right hand sides of the Newman-Penrose equations negligible in comparison with the following ones, as needed in the expansions \eq{5a}-\eq{5f}, one should have $\lambda\ll\kappa\delta t$ and $\mu\ll(\kappa\delta t)^{1/2}$ for $\kappa\delta t \ll 1$ and $\lambda,\mu\ll\kappa\delta t$ for $\kappa\delta t \gg 1$ (in which case we have already assumed the stronger condition $\lambda,\mu\ll1$). Substituting this perturbation into the expressions \eq{6a}-\eq{6d} and \eq{7b}, we can estimate the order of the total relative change in the horizon area during the total quasi-stationary transition for small and large values of $\delta t$ compared to $1/\kappa$,
\ba
&&\frac{\Delta A_{+}}{A_{+}}|_{\kappa\delta t\ll 1}\sim(\Delta_{G}^{2}/(\kappa\delta t)+\Delta_{G}\Delta_{EM}+\Delta^{2}_{EM})\frac{1}{N\kappa\delta t}\,,\non\\
\n{9a}\\
&&\frac{\Delta A_{+}}{A_{+}}|_{\kappa\delta t\gg 1}\sim(\Delta_{G}^{2}+\Delta_{G}\Delta_{EM}+\Delta^{2}_{EM})\frac{1}{N\kappa\delta t}\,.\n{9b}
\ea

Thus, for $A_{+}$ to be an adiabatic invariant, i.e., for $\Delta A_{+}/A_{+}\ll\Delta_{G}$ with respect to the quasi-stationary transition, we must have 
\ba
N|_{\kappa\delta t\ll 1}&\gg&(\Delta_{G}/(\kappa\delta t)+\Delta_{EM}+\Delta^{2}_{EM}/\Delta_{G})\frac{1}{\kappa\delta t}\,,\n{10a}\\
N|_{\kappa\delta t\gg 1}&\gg&(\Delta_{G}+\Delta_{EM}+\Delta^{2}_{EM}/\Delta_{G})\frac{1}{\kappa\delta t}\,.\n{10b}
\ea 
For given values of $\Delta_{G}$ and $\Delta_{EM}$, these expressions show the order of the minimum number of steps for the quasi-stationary transition for $A_{+}$ to be an adiabatic invariant. Note that because the electromagnetic perturbation would be expected to produce a gravitational distortion at least $\Delta_{G}\sim\Delta_{EM}^{2}$, we expect that $\Delta_{EM}^{2}/\Delta_{G}\lesssim1$.

The next step is to consider the fate of the black hole's CH. As it was demonstrated by Poisson and Israel \cite{PI}, as well as by Ori \cite{Ori1,Ori2}, the infinite blueshift of a radiative tail produced at the CH results in the formation of a weak null curvature singularity, the {\em mass inflation null singularity}, at the ingoing part of the CH. In a following investigation by Marolf and Ori \cite{MarOri}, it was shown that for late-infall-time observers (the observers whose Eddington's advanced null coordinate $v_{eh}$ at which they cross the EH, with $v_{eh}=0$ at the formation of the EH, obeys $\kappa v_{eh}\gg 1$), the space-time in the black hole interior may be described by an essentially unperturbed stationary space-time up to very near the CH. At the outgoing portion of the CH the linear gravitational metric perturbations $h_{\mu\nu}$ decouple into a superposition of outgoing and ingoing components and for $\kappa v\gg 1$ the outgoing component decays as an inverse power of $v$. 

The late-infall-time observers who arrive at the outgoing portion of the CH will encounter a {\em gravitational shock-wave singularity}: a finite jump in the metric $\Delta h_{\mu\nu}\sim 1$, which is the shock wave amplitude, within an effectively vanishing observer's proper time (the shock wave width) $\propto e^{-\kappa v_{eh}}\ll1$, where $\kappa$ is the CH surface gravity. All the time-dependent structure of the space-time gets compressed into the shock wave. The shock wave singularity is more violent than the {\em mass inflation null singularity}, and it is detected by both freely falling and accelerated late-infall-time observers. The late-infall-time observers who arrive at the ingoing portion of the CH will encounter the {\em mass inflation null singularity} which decays as an inverse power of $v_{eh}$ and vanishes in the late-infall limit, $v_{eh}\to+\infty$. 

This analysis was done for the Reissner-Nordstr\"om space-time. However, in the papers above arguments were given that the same situation will take place in the case of a rotating Kerr-Newman black hole (see discussions in \cite{Ori2,MarOri}). This result suggests to us to consider, instead of the singular CH, a regular {\em stretched Cauchy horizon}, which is located very near the CH, which keeps track of the perturbation.  A replacement of the CH with the stretched horizon is done by identifying events on the stretched horizon, e.g. an ingoing beam of light, with the corresponding events detected by freely falling observers very shortly before they enter the CH. The stretched horizon area $A_{-}$ is approximately equal to the area of the would-be regular CH. The difference between these areas decreases exponentially with the advanced time $v_{eh}$ after the perturbation and vanishes in the late-infall limit. 

Here we shall assume that the results of Marolf and Ori \cite{MarOri} are valid for a distorted black hole as well as for the undistorted black holes they studied. Then one can define and solve the Newman-Penrose equations \eq{3a}-\eq{3b} at the {\em stretched Cauchy horizon} and take the limit to approach the real Cauchy horizon from the regular interior region and use the equation for the area rate \eq{2}. As a result, $A_{-}$ is an adiabatic invariant as well. Thus, at the end we have both the area $A_{-}$ of the CH and the area $A_{+}$ of the EH remaining adiabatically constant, so that the relation $A_{+}A_{-}=(8\pi J)^{2}+(4\pi Q^{2})^{2}$, which is true for an undistorted Kerr-Newman black hole, remains valid for distorted black holes as well. One can also conclude that not only the area product but any function of both $A_{+}$ and $A_{-}$  remains the same. 

The horizon area product for higher-dimensional black objects was studied as well. For example, Castro and Rodriguez \cite{AM} showed that product of horizon areas is independent of the mass and topology of the horizons for all known 5-dimensional asymptotically flat black rings and black strings. However, as it was illustrated by Visser \cite{Visser}, for different multi-horizon black hole solutions not all area products are mass independent. In the work by Cveti\v c, Gibbons, and Pope \cite{CGP}, explicit results for the product of all horizon areas for general multi-charge black holes were given in terms of quantized charges, angular momenta, and the cosmological constant.\footnote{Note that it was argued by Faraoni and Zambrano Moreno \cite{Far} that quantization rules for the horizon areas of stationary black holes are misleading, for they do not correspond to realistic time-dependent situations.} 

All these solutions are exact, without influence of external matter and fields on the black objects. Thus, it is important to know whether these area products are generic as they are in the case of the Kerr-Newman black hole. Therefore, it is natural to ask whether our heuristic argument can be generalized to higher-dimensional black objects. We can follow the same scenario and show that, assuming a black object is stable against an external perturbation due to a (higher-dimensional) weak gravitational wave generated by motion of distorting matter which preserves some symmetry (an analogue of the axial symmetry) such that the black object's angular momenta (as well as charges) are preserved, the outer horizon area does not change due to an adiabatic perturbation. As far as the inner horizons, some arguments were given in \cite{MarOri} that the late-time limit of black hole interiors can be carried over to generic stable black holes in any dimension. This implies that we can repeat our scenario by introducing the stretched inner horizon(s), as we did in the case of the Kerr-Newman black hole, and conclude that the inner horizon area(s) do not change either. 

Thus, we can formulate the following\\ 
{\bf Conjecture}: {\em Assume one has a d-dimensional $(d\geq4)$, stationary, multi-horizon black-object solution with the horizon areas $\{A_{i},i=1,...,n\}$, a set of charges $Q_{j}$, and angular momenta $\{J_{k},k=1,...,[(d-1)/2]\}$, and possibly with a cosmological constant $\Lambda$, such that the following relation holds: 
\be
\prod_{i=1}^{n}A_{i}=f(Q_{j},J_{k},\Lambda^{-1/2})\,,\non
\ee
where $f$ is some polynomial function. Then, assuming that the solution is stable against external gravitational perturbations, this relation will hold for any adiabatically distorted solution with the same values of the charges and angular momenta.}
 
As an example illustrating this conjecture, we can consider a five-dimensional, static, electrically charged black hole distorted by an external, static and electrically neutral distribution of matter. Then the product of the areas of the outer and inner horizons is proportional to the cube of the electric charge \cite{AS,SA}. 

One of the possible issues related to this conjecture is a consideration of the inner horizons of a multi-horizon solution which are located behind the outermost inner one (assuming that the solution has nested inner horizon topology.) Due to the gravitational perturbation of the outermost inner horizon, the mass inflation phenomenon and formation of a gravitational shock wave singularity will seal off the interior region. As a result, only two areas (of the outer horizon and the outermost inner horizon) can be involved in the construction of the area product. However, one may consider instead of the perturbed solution an eternal one, just as it was done in the works by Ansorg and Hennig. In such a consideration, one can formally ignore the ``sealing issue'' and address each inner horizon independently. The validity of this approach, as well as verification of the proposed conjecture, is an open issue. 

Let us summarize our results. In this paper we have proposed an heuristic argument for the universal area relation of a four-dimensional adiabatically distorted Kerr-Newman black hole. Based on this argument we formulated the conjecture for an adiabatically distorted multi-horizon black object whose product of horizon areas is expressed in terms of a polynomial function of its angular momenta, charges, and the inverse square root of the cosmological constant. This conjecture addresses the question by Cveti\v c, Gibbons, and Pope \cite{CGP} about the quantization of the product of horizon areas of a black object in the presence of external fields, and it may be useful for study of the microscopic properties of such black objects in terms of field theories in more than two dimensions.   


The authors are grateful to the Natural Sciences and Engineering Research Council of Canada for its support.  We also appreciate comments by an anonymous referee.



\end{document}